# MindBenchAI: An Actionable Platform to Evaluate the Profile and Performance of Large Language Models in a Mental Healthcare Context

Bridget Dwyer[1]*, Matthew Flathers[1]*, Akane Sano[2], Allison Dempsey[3], Andrea Cipriani[4], Asim H. Gazi[5], Carla Gorban[6], Carolyn I. Rodriguez[7], Charles Stromeyer IV[8], Darlene King[9], Eden Rozenblit[1], Gillian Strudwick[10], Jake Linardon[11], Jiaee Cheong[1], Joseph Firth[12], Julian Herpertz[13], Julian Schwarz[14], Margaret Emerson[15], Martin P. Paulus[16], Michelle Patriquin[17], Yining Hua[18], Soumya Choudhary[19], Steven Siddals[1], Laura Ospina Pinillos[20], Jason Bantjes[21], Steven Scheuller[22], Xuhai Xu[23], Ken Duckworth[24], Daniel H. Gillison[24], Michael Wood[24], John Torous[1]**

1. Division of Digital Psychiatry, Beth Israel Deaconess Medical Center, Harvard Medical School, Boston, MA, USA.
2. Department of Electrical and Computer Engineering, Rice University, Houston, TX, USA.
3. Department of Psychiatry and Family Medicine, School of Medicine, Centers for American Indian and Alaska Native Health, Colorado School of Public Health, Telemedicine Helen and Arthur E. Johnson Depression Center, University of Colorado Anschutz Medical Campus, Aurora, CO, USA.
4. Department of Psychiatry, University of Oxford, UK; Oxford Precision Psychiatry Lab, National Institute for Health and Care Research (NIHR) Oxford Health Biomedical Research Centre, Oxford, UK.
5. School of Engineering and Applied Sciences, Harvard University, Cambridge, Massachusetts, USA.
6. The Brain and Mind Centre, The University of Sydney, Camperdown, New South Wales, Australia.
7. Department of Psychiatry and Behavioral Sciences, Stanford University, Stanford, CA, USA.
8. Peer Advisory Advocacy and Research Council, Massachusetts Mental Health Center Public Psychiatry Division of the Beth Israel Deaconess Medical Center, Boston, MA, USA.
9. Department of Psychiatry, The University of Texas Southwestern Medical Center, Dallas, TX, USA.
10. Centre for Addiction and Mental Health, Toronto, Ontario, Canada.
11. School of Psychology, Deakin University, Geelong, VIC, Australia.
12. Division of Psychology and Mental Health, University of Manchester, Manchester Academic Health Science Centre, Manchester, UK.
13. Department of Psychiatry and Neuroscience, Campus Benjamin Franklin, Charité–Universitätsmedizin Berlin, Berlin, Germany
14. Department of Psychiatry and Psychotherapy, Center for Mental Health, Immanuel Hospital Rüdersdorf, Brandenburg Medical School Theodor Fontane, Rüdersdorf, Germany.
15. College of Nursing, University of Nebraska Medical Center, Omaha, USA.
16. Laureate Institute for Brain Research, Tulsa, OK USA.
17. Baylor College of Medicine, One Baylor Plaza, Houston, Texas, USA.

18. Department of Epidemiology, Harvard T.H. Chan School of Public Health, Boston, MA, USA
19. Department of Psychiatry, National Institute of Mental Health and Neurosciences, Bangalore, India.
20. Institute for Life Course Health Research, Department of Global Health, Faculty of Medicine and Health Sciences, Stellenbosch University, Stellenbosch, South Africa
21. Department of Psychiatry and Mental Health, Faculty of Medicine, Pontificia Universidad Javeriana, Bogota, Colombia
22. Department of Psychological Science, University of California, Irvine, CA, USA.
23. Department of Biomedical Informatics, Columbia University, NY, USA.
24. National Alliance on Mental Illness (NAMI), Arlington, Virginia.

*Authors contributed equally
**Corresponding author. Email: jtorous@bidmc.harvard.edu

## Abstract

Individuals are increasingly utilizing large language model (LLM)-based tools for mental health guidance and crisis support in place of human experts. While AI technology has great potential to improve health outcomes, insufficient empirical evidence exists to suggest that AI technology can be deployed as a clinical replacement; thus, there is an urgent need to assess and regulate such tools. Regulatory efforts have been made and multiple evaluation frameworks have been proposed, however,field-wide assessment metrics have yet to be formally integrated. In this paper, we introduce a comprehensive online platform that aggregates evaluation approaches and serves as a dynamic online resource to simplify LLM and LLM-based tool assessment: *MindBenchAI*. At its core, *MindBenchAI* is designed to provide easily accessible/interpretable information for diverse stakeholders (patients, clinicians, developers, regulators, etc.). To create *MindBenchAI*, we built off our work developing MINDapps.org to support informed decision-making around smartphone app use for mental health, and expanded the technical MINDapps.org framework to encompass novel large language model (LLM) functionalities through benchmarking approaches. The *MindBenchAI* platform is designed as a partnership with the National Alliance on Mental Illness (NAMI) to provide assessment tools that systematically evaluate LLMs and LLM-based tools with objective and transparent criteria from a healthcare standpoint, assessing both profile (i.e. technical features, privacy protections, and conversational style) and performance characteristics (i.e. clinical reasoning skills). With infrastructure designed to scale through community and expert contributions, along with adapting to technological advances, this platform establishes a critical foundation for the dynamic, empirical evaluation of LLM-based mental health tools—transforming assessment into a living, continuously evolving resource rather than a static snapshot.

**Introduction**

The rapid rise of artificial intelligence (AI), fueled by the new generation of large language models (LLMs), has been highly visible in the mental health space [1]. LLMs are generative AI systems trained on vast amounts of text data that generate human-like responses by predicting contextually likely word sequences– thus creating conversational experiences that millions now access through interfaces like ChatGPT, Claude, Gemini and other emerging specialized mental health applications. A Harvard Business Review survey noted that mental health may be the single highest use case of this technology [2]. Survey research suggests that more than 30% of people may already seek emotional support from LLMs [ 3, 4], and real-world clinical experiences further demonstrate that the use of LLMs in mental health constitutes a dynamic and rapidly advancing field of significant importance [5, 6].

But this rapid rise has also raised many questions and concerns [7]. There is undeniable evidence that LLMs can cause harm, ranging from incorrect medical advice to overreliance [8-11]. Some people have even exhibited signs of emotional dependence on LLMs, and the conversational dynamics/overuse of some LLMs have led to harmful parasocial relationships [12, 13] and even induced what is currently called "AI Psychosis" or "cognitive pattern amplification," [14, 15] where delusions are co-created with the LLM. Weekly news stories share tragic cases where LLM use has been linked with suicide [16, 17], and the safety of AI for mental health is now a focus not only of regulators but also of the AI companies themselves [18, 19]. Yet the regulation and safety assessment of LLMs and LLM-based tools is not simple as these models can respond to a near-infinite variety of prompts and produce a near-infinite variety of outputs, with the same prompt leading to both desirable and undesirable responses in different instances due to the probabilistic nature of LLMs. In response, at least 60 AI evaluation/regulatory frameworks have been published to date [20-82]. While each offers unique merits, none of them can provide actionable information to guide safer evaluation of LLMs today.

Formal regulatory pathways for LLMs in mental health, even when eventually developed, may not offer clinicians and patients the information needed to make informed choices. The vast majority of existing mental health smartphone apps declared themselves to be wellness tools and so were exempt from any regulation [83]. The same is already happening in the mental health LLM space, with most companies offering 'therapy' but including disclaimers in the fine print noting they are not offering clinical services and so will not follow clinical regulations [84]. This trend presents an urgent challenge. Despite the rapid expansion of LLM use in mental health and the proliferation of tools marketed directly to patients, there is a striking lack of empirical evidence or standardized guidance to help patients and clinicians evaluate the safety, effectiveness, and potential risks of any given LLM. Without such evidence, both clinical decision-making and patient trust remain vulnerable, underscoring the need for systematic evaluation frameworks and rigorous research.

In response, our team has proposed an actionable solution that expands on the approach previously utilized to address the same challenge around mental health app evaluation. For the last ten years, we have supported MINDapps.org as the largest mental health app database, providing free access to anyone in the world to transparent, actionable, and updated information on mental health apps [85, 86]. Through evaluating apps across 105 dimensions, MINDapps.org allows our team to share a profile for each app that a user can then search to assess which apps offer the best match for their unique needs [87]. For MINDapps.org, we profile each app every 6 months and have designed the 105 profile questions to assess the core features, inputs, and outputs that most apps can offer [88].

However, with LLMs, a profiling approach like MINDapps.org is useful, but alone, insufficient. While certain profiling aspects, such as privacy/security, are relevant in both apps and AI, others, such as personality, are unique to AI agents. Given that traditional smartphone apps are programs that do not often perform reasoning but instead support skills in a deterministic manner (e.g., most of these apps offer mindfulness or mood tracking [88]), assessing their performance was superfluous. However, with LLMs, assessing performance is critical as their abilities to respond are neither deterministic nor limited to supporting skills.

Benchmarks are the most widely used ways to assess LLM performance [89, 90]. When new statistics emerge and claim that a certain LLM is "better" than another, this generally refers to benchmarks or standardized tests that are used to assess LLM performance. While general benchmarks for language understanding, mathematical reasoning, and coding proliferate, mental health evaluation remains fragmented and poorly suited to clinical realities [91]. Existing medical LLM benchmarks either exclude mental health entirely or treat it as a minor component– MedQA includes only 5-15% psychiatry questions [92], while broader efforts like MedHELM [72] and GMAI-MMBench [78] span dozens of medical specialties with minimal mental health representation. In May of 2025, OpenAI shared HealthBench, a set of AI-generated cases rated by clinicians [49]. OpenAI shared this benchmark to enable open LLM benchmarking and ascertain how LLM-derived responses compared to the clinician assessments. While HealthBench is an important milestone, only a few of the cases focused on mental health with even fewer mental health experts scoring those cases (ie, most were scored by non-mental health clinicians), thus calling into question its reliability to serve as an appropriate mental health benchmark.

Even dedicated mental health benchmarking attempts reveal significant gaps. Recent domain-specific efforts like PsychBench targeting psychiatric clinical tasks with structured scoring [93], MentalChat16K providing conversational data for testing assistants [94], and CounselBench offering clinician-rated adversarial counseling evaluations [95] all advance the field, but they employ incompatible formats, task designs, and scoring rubrics that hinder meaningful cross-benchmark comparisons. Most critically, no centralized platform exists for *accessing* and *comparing* results *across evaluation efforts*, leaving stakeholders unable to make informed comparisons between models even though millions are already using LLM-based tools for mental health support.

In this paper, we outline how a new approach, combining profiling with performance metrics, is now feasible and offers real-time, actionable information for evaluating LLMs and LLM-based tools in mental health contexts. This method also aids researchers in assessing the potential of current models, regulators in identifying safety risks, and companies in rapidly improving their models to deliver better outcomes.

## Methods

Building on our experience from MINDapps.org, and a decade of work in the health technology evaluation space [85, 86], we developed a comprehensive approach to LLM evaluation that offers both profiling and performance evaluation. We then combined both approaches into a single dashboard designed to share real-time, transparent, and actionable information for all stakeholders. Given the expanding role of LLMs for patients and families, we partnered with the National Alliance on Mental Illness (NAMI), the largest grassroots mental health organization in the United States, to ensure this evaluation approach reflects the expertise and needs of individuals and families affected by mental health conditions.

**Profile Evaluation**
In alignment with the American Psychiatric Association's (APA) app evaluation model (and its AI corollary), our LLM profiling methodology addresses core factors such as data use, privacy, and interactional dynamics relevant to any LLM–regardless of its clinical performance [97, 98]. An LLM that offers superior mental health support but *owns* a user's personal health information presents an individual choice that users can only make if profiling information is accessible [99]. Through profile evaluation, critical information that is often challenging for patients or clinicians to verify each time they use an LLM or LLM-based tool is clearly presented and feasibly navigated.

*Technical Profile*

While LLMs and LLM-based tools are more complex than apps, they remain technological systems that require a systematic assessment of technical properties, which directly impact user safety and privacy. To identify which technical characteristics warranted inclusion in our platform, three independent raters systematically reviewed the full 105 questions from MINDapps.org to determine individual relevance for LLM evaluation. When disagreements arose, group discussion continued until consensus was reached. We then sought public feedback through public webinars with the Society of Digital Psychiatry, current MINDapps.org users, and our team's patient advisory board. 48 of the original 105 questions from MINDapps.org were retained in the final curated list, which includes universal digital health concerns such as data retention, privacy policies, and developer information. Through this process, we also identified 59 new characteristics specific to LLM deployment that required documentation, including token limits, context window specifications, model versioning practices, API reliability guarantees, and conversation memory management. For a list of all profiling questions, please see [Appendix A](#).

Following the MINDapps framework's emphasis on objectivity, we structured each characteristic to produce either a binary (yes/no) or numerical answer, enabling systematic comparison across tools. Through our review, we determined that questions needed to be organized into two distinct categories: base model characteristics (training data transparency, security certifications, API limitations) and tool-specific implementations (conversation storage policies, user authentication methods, content filtering approaches). This dual-level structure emerged from recognizing that users may interact with both the underlying model and its specific implementations, each introducing distinct privacy and safety considerations.

*Conversational Dynamics Profile*

User feedback and emergent research informed our approach to profiling LLM conversational dynamics that, cumulatively, users may perceive as personality. While smartphone apps are seldom associated with personalities, many people may anthropomorphize LLMs [100]. When ChatGPT-5 was released with less personality, users were unhappy, so the company reinstated LLM personification [101]. Concerns have arisen around LLMs being too sycophantic/agreeable, and causing harm by inadvertently supporting delusions/harmful ideas [102]. Thus, the importance of understanding LLM conversational limits cannot be understated. Standardized personality assessments unify natural-language and trait concepts into singular frameworks, providing a descriptive classification system to categorize default LLM interaction styles [103]. We investigated the International Personality Item Pool [104] to identify which frameworks could be feasibly adapted for LLM assessment. Selection criteria included: validated psychometric properties, ability to be reformulated as prompts, relevance to therapeutic interactions, and interpretability by non-specialist audiences.

**Performance Evaluation**

Beyond understanding how LLMs are deployed and experienced through profiling, performance metrics are necessary for assessing the capabilities, benefits, and harms. Based on the most recent research on LLM performance evaluation, we sought to capture performance evaluation not only via benchmark scores (the models' conclusions) but also via reasoning analysis (how the model arrives at those conclusions) [105]. The latter is important, as models that arrive at correct answers for incorrect reasons are likely to perform well on benchmarks, but fail when confronted with novel cases or clinical situations [106].

*Benchmarking*

To determine optimal benchmark formats for mental health LLM evaluation, we systematically analyzed existing assessment approaches across clinical psychology, ML evaluation, and medical education. We identified that traditional binary correctness metrics used in medical benchmarks like MedQA (based on questions related to the US National Medical Board Examination (USMLE)) [107] failed to capture the clinical reality of mental health care, where there is not a single yes/no answer and depth of the field beyond medical student exam questions. As noted above, OpenAI's HealthBench is another benchmark but not appropriate for mental health use cases as the clinical scenario data is not specific to mental health interactions.

To identify a scalable standard format for our mental health benchmarks, we further reviewed existing clinical scales and model benchmarks against several criteria: ability to cover diverse mental health domains (crisis intervention, psychopharmacology, therapeutic boundaries), production of numeric outputs enabling quantitative comparison, straightforward administration to language models through text prompts, capacity to capture reasoning alongside answers, and compatibility with expert validation processes. These requirements reflect the need to move beyond binary yes/no responses to capture the nuanced judgments characteristic of mental health practice, while remaining structured enough for systematic, repeatable, and consistent evaluation and scoring. While no established system meets all these requirements, we identified the format used in cases related to safety evaluation through the Suicide Intervention Response Inventory 2 (SIRI-2) assessments, as administered to LLMs by McBain et al. [108], as meeting our criteria.

This format offered multiple advantages. Numeric expert ratings preserve information about degrees of appropriateness that binary metrics would lose. The distribution of expert scores allows for both consensus and legitimate disagreement, valuable for understanding where professional judgment itself remains unsettled. The data structure itself generates preference pair data compatible with supervised fine-tuning and reinforcement learning approaches, enabling AI developers to use evaluation data for model improvement. And most importantly, this format's flexibility allows diverse mental health assessments to maintain their domain-specific focus, whether questions about substance abuse or schizophrenia, while sharing a consistent data structure and even enabling detailed assessment of each interaction/response, as shown below in Figure 1.

```
client:"But my thoughts have been so terrible... I could never tell them to anybody."
helper_a:"You can tell me. I'm a professional, and have been trained to be objective about these things." (expert mean score: -2.14, SD of expert scores: 1.07)
helper_b:"So some of your ideas seem so frightening to you, that you imagine other people would be shocked to know you are thinking such things." (expert mean score: 2.14, SD of expert scores: 0.38)
```
**Figure 1**: Question 3 of the SIRI-2 Scale

The same format can also support clinical case analysis, where patient scenarios are paired with potential clinical decisions, as shown below in Figure 2:

```
Case: Patient A is a 25-year-old white male who has never been psychiatrically
hospitalized and has a history of depression in the late teens. He was trialled on
Zoloft for 2 weeks but was discontinued for unclear reasons. He presents today with
decreased sleep, high energy, and family members concerned that he's talking fast. He
has been staying up late at night developing a new code.
Decision_a: The clinician asks for further details about family history of bipolar
disorder or any substance use.
Decision_b: The clinician prescribes trazodone for poor sleep.
```
**Figure 2**: Question 1 of the Adversarial Psychopharmacology Case Scale currently undergoing validation

This unified format enables benchmarking across different mental health domains to coexist within the same evaluation infrastructure, maintaining comparability across all assessments.

**Reasoning Analysis**

We investigated methods for systematically analyzing model decision-making processes to understand how LLMs arrive at conclusions/responses. We reviewed approaches from clinical education (case-based reasoning assessment), AI interpretability research (chain-of-thought prompting, mechanistic interpretability), and adversarial testing literature. We sought a reasoning assessment method that included: feasibility for non-technical evaluators, scalability across thousands of responses, ability to identify systematic patterns rather than isolated errors, and compatibility with our benchmark format.

Chain-of-thought prompting emerged as the most implementable approach, requiring only modification of our benchmark prompts to request step-by-step reasoning before numerical ratings. To analyze reasoning outputs at scale, we developed a multi-method approach combining natural language processing and text embedding analysis. Given the importance of assessing reasoning beyond the retrieval of facts, we identified key adversarial techniques particularly relevant to mental health, including but not limited to: information gaps, counterfactual variations, distractor information, and anchoring bias tests. By documenting which techniques are present in each benchmark item's metadata, we can automatically analyze whether models' reasoning may fail under specific types of challenges.

This resulting infrastructure, designed to support profiling and performance (accuracy and reasoning), was built for extendibility and sharing, so other teams can easily add assessment questions. An advantage of this approach is that the benchmarks can be scheduled and run automatically to transparently evaluate a host of LLMs, ensuring all stakeholders have access to the information they need to make more informed decisions.

## Results

We developed a web-based platform, *MindBenchAI*, for LLM evaluation through integrated profiling and performance assessment. The interface organizes evaluation data through four primary components: (1) technical profiles, (2) conversational dynamics, (3) benchmark leaderboards, and (4) reasoning analyses. See figure 3. Each component addresses specific stakeholder needs identified during our methods development while maintaining interoperability that enables comprehensive evaluation across multiple dimensions.

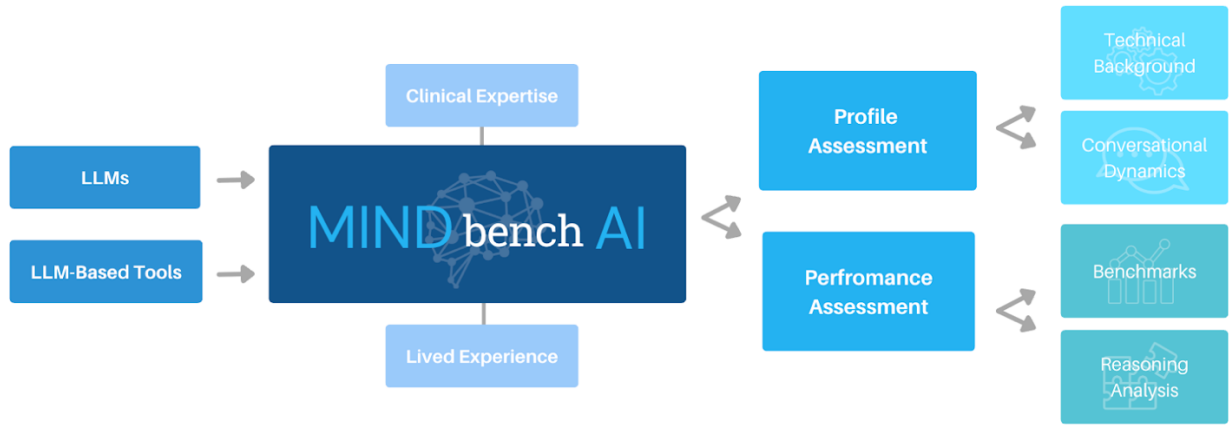

**Figure 3:** MNDbenchAI Evaluation Logic

*(1) Technical Profile*

The technical profile interface (see figure 4 below) displays information through a MINDapps.org inspired display with dual-tabbed views separating base models from downstream tools. Each profile presents responses in a structured format with binary indicators using checkmarks for immediate visual scanning, supplemented by expandable sections containing detailed explanations and verification sources. The dashboard prioritizes critical safety information through a hierarchical display. Each data point links to its verification source, whether from official documentation, terms of service, or direct testing, with timestamps indicating when information was last verified. We also created a comparison tool that allows for the side-by-side examination of up to three models or tools, with differences highlighted in contrasting colors. Export functionality generates PDF reports suitable for institutional review boards or compliance committees evaluating potential LLM deployments.

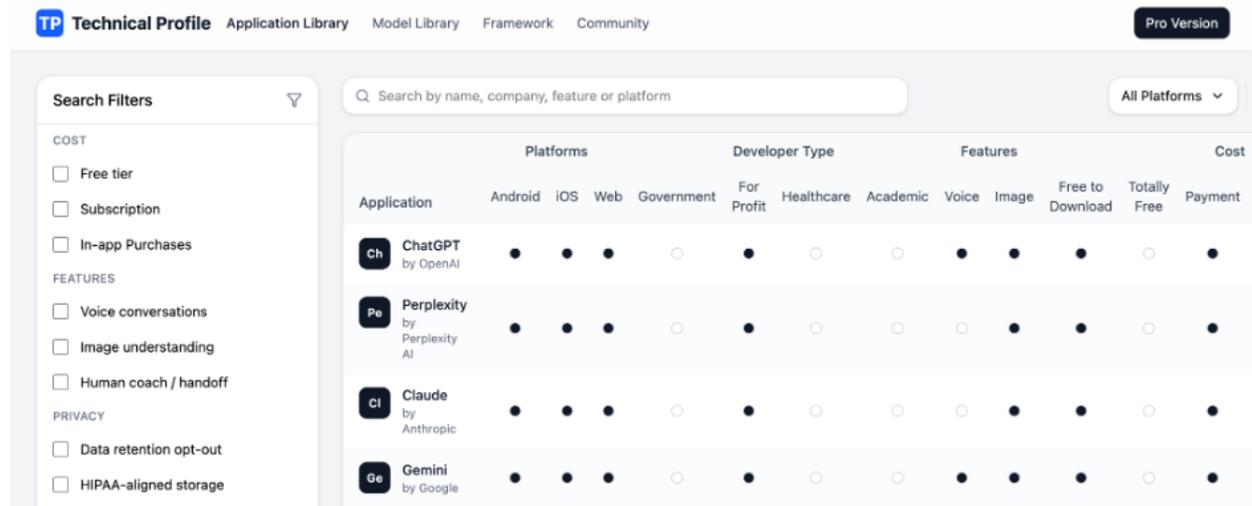

**Figure 4:** Technical Profile section of our platform

*(2) Conversation Dynamics Profile*

The conversational dynamics profile interface (see Figure 5 below) displays results through framework-specific visualizations of different personality assessments. Big Five and HEXACO results appear as radar charts showing dimensional scores, Myers-Briggs Type Indicator (MBTI) displays as a four-letter type with percentage strengths for each dimension, and Enneagram shows as a primary type as a number from one to nine, with alternative dimensions as percentages. Users can toggle between frameworks using tabs, with each visualization optimized for its assessment type. Benchmark ranges derived from [human therapist norms/general population data] appear as reference points, enabling users to contextualize whether a model exhibits unusual personality patterns.

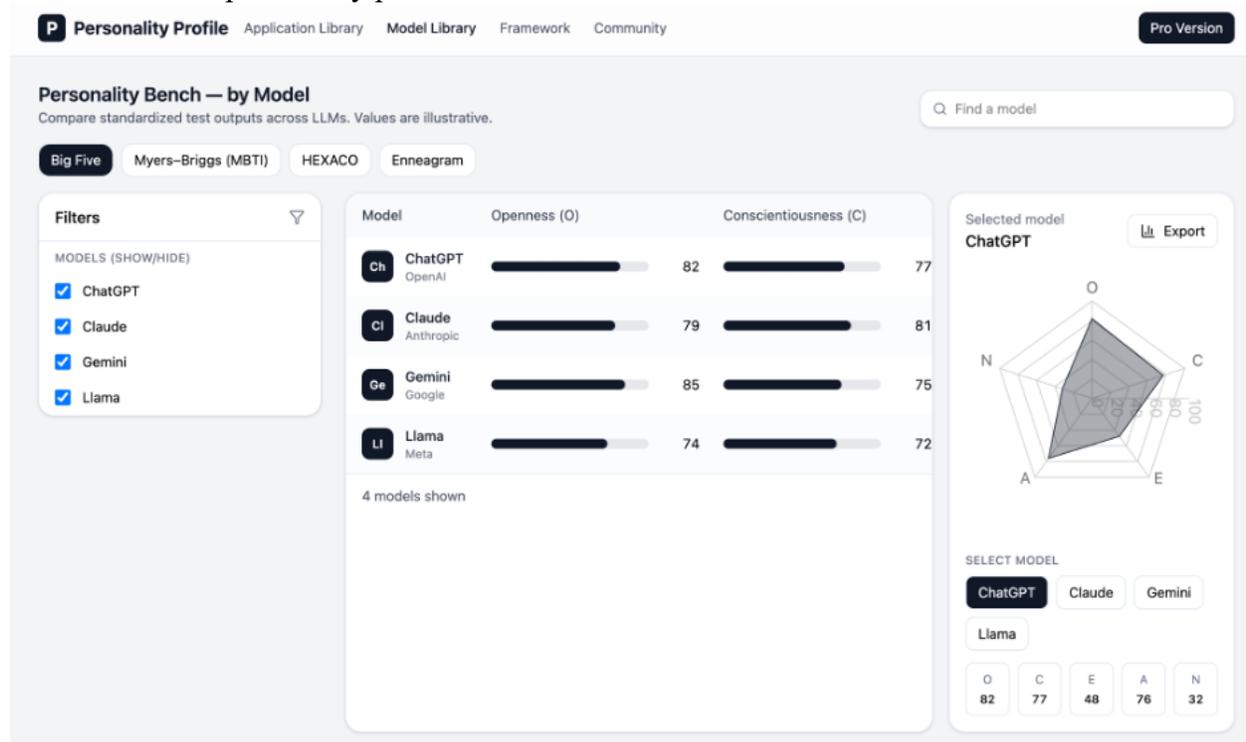

**Figure 5**: Conversational Dynamics section of our platform assessed via Personality Tests

*(3) Benchmarks*

From our review of existing clinical scales and mental health benchmarks, we identified those that met our format requirements of prompts paired with expert-rated responses. To expand coverage across additional mental health domains, we collaborated with psychiatry residents at BIDMC to develop an additional suite of 75 clinical case benchmarks covering psychopharmacology, peri-natal mental health, and psychiatric diagnosis, which serve as the initial evaluation instruments as the platform undergoes beta testing.

We chose to display results through a leaderboard format (see Figure 6 below), adapting the established convention from machine learning evaluation platforms like HuggingFace and

OpenLLM Leaderboard that developers already understand while ensuring the interface remains interpretable for clinical users unfamiliar with technical benchmarking. The platform maintains separate leaderboard tabs for base models (e.g. GPT-5, Claude Opus 4, Gemini 2.5 Pro) and downstream tools (e.g. Character.AI or other specialized therapy bots building off these base models). Each row in the leaderboard represents a model or tool, while the columns display performance across different validated benchmarks: SIRI-2 for crisis response, as well as clinical case benchmarks for psychopharmacology and perinatal mental health (currently undergoing validation). The infrastructure is designed to seamlessly integrate new benchmark cases/questions/scenarios as they are uncovered and developed.

Similar to MINDapps.org, rather than combining scores into a single overall ranking that would obscure essential differences [109], we preserve granular performance data, allowing users to identify models that excel in specific domains.

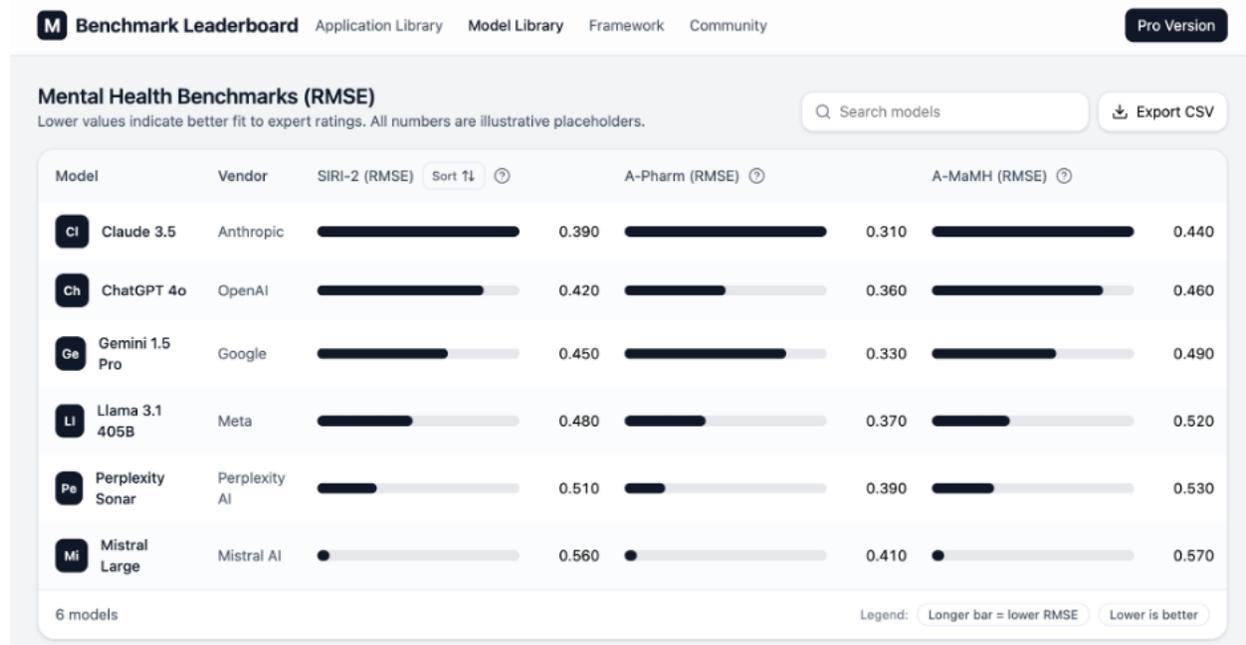

**Figure 6:** Mental Health Benchmark Leaderboard on our platform.
*RMSE = Root Mean Square Error, a measure of accuracy

*(4) Reasoning Analysis*

In order to operationalize reasoning assessment beyond benchmark scores, our platform implements chain-of-thought extraction for every benchmark item, prompting models to articulate their step-by-step reasoning before providing numerical ratings. To systematically probe reasoning robustness, we encouraged benchmark developers (ie, those writing cases and questions) to incorporate specific adversarial techniques (see Appendix B) into their scenarios, as noted above. Those techniques are documented in question metadata, thus enabling analysis of which reasoning challenges might cause systematic failures.

The reasoning interface (see Figure 7 below) provides granular analysis at multiple levels. Each benchmark's dedicated page displays a performance breakdown by individual questions, thus revealing item-specific patterns that could indicate particular reasoning vulnerabilities.

The platform's reasoning infrastructure is designed to prioritize extensibility based on emerging technical methods and community feedback. Our initial implementation– COT extraction and adversarial technique tracking– provides immediate value whilst establishing the technical architecture for future enhancements as the platform matures.

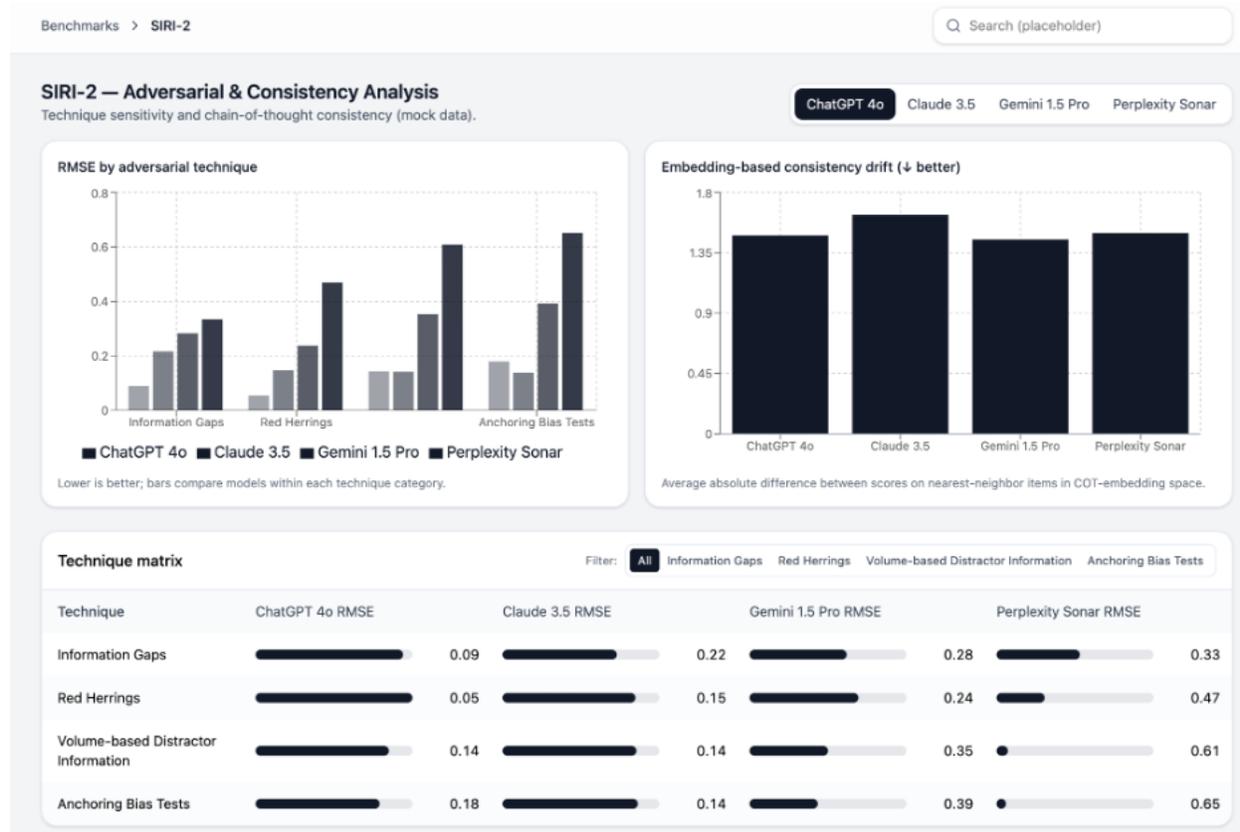

**Figure 7:** Benchmark Specific Reasoning Analysis page

## Discussion

This paper introduces a scalable and transparent means to assess LLMs and LLM-based tools, inspired by the MINDapps.org approach for smartphone app evaluation, but adding specific profiling and performance features unique to LLMs. The resulting *MindBenchAI* platform is designed to be easy for others to add their own benchmarks, as the platform can easily scale through partnerships and collaborations with patients, clinicians, and organizations. Anchored by a partnership with NAMI, the platform is designed from the ground up to ensure it meets the needs of individuals and families affected by mental health conditions and grows with the expertise and values that NAMI embodies. By enabling all parties to transform their values and evaluation needs into actionable results that can be broadly shared and easily accessed,

*MindBenchAI* offers a pragmatic and evidence-based solution to LLM evaluation in mental health.

*MindBenchAI* is informed by our decade of experience in evaluating mental health apps, reviewing over 60 current health AI evaluation models, as well as numerous reviews of the AI mental health space [110-113], and discussions with patients, clinicians, regulators, and developers. While the true value of this system will be evidenced through its utilization, the approach is designed to align with the needs of all stakeholders. It does not require companies to disclose sensitive information (OpenAI already created its own benchmark), it avoids making regulatory claims that require legislation to enact, and it is easy for patients and clinicians to access without any technical knowledge. Finally, *MindBenchAI* does not conflict with any of the AI frameworks reviewed [20-82] as those questions and ideas can be added as benchmarks that the platform can host and help to disseminate.

Beyond assisting LLM evaluation, an additional benefit of this approach is the support and guidance it offers developers to ensure their generative models improve. Expert ratings generate preference pairs that developers can use for model training to help them release models that are more effective in mental health contexts. Identified failures become specific targets for enhancement that can be fixed before a model is publicly released. As models improve and new edge cases emerge, the community can develop new cases (benchmarks) to focus on emerging failure modes or expand into new assessment domains that the models should be evaluated on. The platform's architecture accommodates future advances in interpretability techniques and regulatory frameworks without requiring fundamental restructuring, thus ensuring that infrastructure investments made today remain valuable even as the underlying technology rapidly evolves.

Like any evaluation system, benchmarks are only as good as the involvement of the domain experts who develop specialized cases for their fields. The *MindBenchAI* framework provides the technical scaffolding and infrastructure, and already supports over 100 cases, but it will be more useful with the input and collaboration from more clinical experts and NAMI members. Developers can contribute by enhancing our open-source codebase, sharing internal safety benchmarks, providing API credits for evaluation, or using our preference pair data to fine-tune their models. Researchers can leverage the aggregated data for studies–such as cross-cultural comparisons and longitudinal safety analyses–that would be impossible with isolated evaluations. We particularly encourage contributions from underrepresented perspectives in mental health, as benchmarks developed primarily by Western, English-speaking clinicians often embed cultural assumptions that limit global applicability and risk inequitable outcomes. This platform is designed to evolve infrastructure for multilingual benchmarks and cultural adaptation tracking to support scalability, although realizing this potential requires active community contribution. We invite mental health organizations, academic institutions, and community advocates to partner with *MindBenchAI* to co-create benchmarks, participate in governance, and ensure the platform reflects diverse global needs while prioritizing ethical and safe deployment and application.

A key feature of *MindBenchAI* is its "living mode" which will keep the ever-expanding platform up to date and ready to support real-world implementation of LLM models. Leveraging existing projects in the space of evidence synthesis and living systematic reviews [114], we have the experience and technical knowledge to accelerate AI innovation from discovery science into effective new tools for mental health disorders, which will inform regulators, industry and clinical practice across the world.

While MINDbenchAI represents an urgently needed step toward systematic evaluation, it is also clear that the field of AI for mental health remains pre-paradigmatic in several critical respects. First, we lack a shared framework of common usage patterns. In practice, "using AI for mental health" encompasses a wide range of behaviors, from seeking companionship or advice, to crisis support, trauma work, journaling analysis, role play, or structured delivery of therapeutic modalities such as CBT or DBT [4]. Each of these patterns may carry distinct risk–benefit profiles and require different evaluation approaches. This also raises a deeper question of what we mean by an "AI tool" in this domain. While our present analysis distinguishes between base models and apps built upon them, in lived use the true differentiator for safety and effectiveness may often lie in what the user asks for (the usage pattern) and how they frame it (the prompt). Prompts can override default personalities [115], app-level system instructions, and even aspects of model performance [116]. A truly comprehensive evaluation library may therefore need to include not only models and apps, but also representative prompts and scenarios that capture these usage patterns.

Second, there is a need for benchmarking paradigms that capture patient impact. Current benchmarking approaches, such as comparing AI ratings of single interaction pairs with expert judgments of appropriateness [108] or using LLMs to rate multi-turn conversations against human-centric criteria [117], are an important start, but ultimately, we need to evaluate how sustained use of these tools over time affects user wellbeing and clinical outcomes. With a partnership with NAMI, we hope to co-develop these needs solutions.

## Conclusion

This platform represents a necessary step toward empirical evaluation of LLM-based mental health tools that millions already use without systematic mental health impact assessments. By combining comprehensive profiling of deployment contexts and conversational "personality" characteristics with standardized performance measurement through benchmarks and reasoning analysis, we provide stakeholders with actionable intelligence about how these systems actually behave in mental health contexts. The infrastructure is built to scale through community contribution and designed to evolve with technological advances, creating a living resource rather than a static snapshot. While no evaluation framework can eliminate all risks from generative models in mental health, establishing shared empirical foundations for understanding these tools' capabilities and limitations is essential for moving from reactive responses to harmful incidents toward proactive safety improvement. We invite clinical experts by training or lived experience, family members, developers, researchers, and other stakeholders to join us and NAMI to contribute their expertise through benchmark development, model evaluation, and platform enhancement, collectively building the evidence base necessary for responsible AI deployment in mental health.

**Disclosures:**
In the last 3 years, CIR has served as a consultant for Biohaven Pharmaceuticals, Osmind, and Biogen; and receives research grant support from Biohaven Pharmaceuticals, a stipend from American Psychiatric Association Publishing for her role as Deputy Editor at The American Journal of Psychiatry, a stipend for her role as Deputy Editor at Neuropsychopharmacology, and book royalties from American Psychiatric Association Publishing. JF is supported by a UK Research and Innovation Future Leaders Fellowship (MR/Y033876/1) and the NIHR Manchester Biomedical Research Centre (NIHR203308). The views expressed are those of the author(s) and not necessarily those of the NIHR or the Department of Health and Social Care. JF has provided consultancy, speaking engagements and/or advisory services to Atheneum, Bayer, ParachuteBH Ltd, LLMental, Nestle UK, HedoniaUS and Arthur D Little, independent of this work. JT is a clinical adviser for Boeringer Ingelheim. AS has been a consultant for Suntory Global Innovation Center. AS received honoraria from Oak Ridge Associated Universities, Nara Advanced Institute of Science and Technology, Taiwanese Society for Nutritional Psychiatry Research, Korea Advanced Institute of Science and Technology, Amrita Vishwa Vidyapeetham, European Science Foundation, National Science Foundation, Dartmouth College, and has travel support from Apple and Taiwanese Society for Nutritional Psychiatry Research. AS received research funding from Meta, General Motors, Sony, POLA, and NEC.

The other authors report no additional financial or other relationships relevant to the subject of this manuscript.

**Acknowledgments:**
National Institutes of Health (K99EB037411). Andrea Cipriani is supported by the National Institute for Health Research (NIHR) Oxford Cognitive Health Clinical Research Facility, by an NIHR Research Professorship (grant RP-2017-08-ST2-006), by the NIHR Oxford and Thames Valley Applied Research Collaboration, by the NIHR Oxford Health Biomedical Research Centre (grant NIHR203316) and by the Wellcome Trust (GALENOS Project). The views expressed are those of the authors and not necessarily those of the UK National Health Service, the NIHR, or the UK Department of Health and Social Care. JT is supported by the Argosy Foundation and Shifting Gears Foundation.